\documentclass[english,aps]{revtex4}
\usepackage{graphicx}
\usepackage{amsmath}

\catcode`\@=11
\begin{document}

\title{A new multi-component CKP hierarchy
\footnote{ Corresponding author:+8610 68916895}\footnote{E-mail
address:wuhongxia@bit.edu.cn}}

\author{Hongxia Wu}
\affiliation{1. Department of Mathematics, Jimei University, Xiamen,
361021, China \\ 2. Department of Mathematics, Beijing Institute of
Technology, Beijing 100081,China }
\author{Xiaojun Liu}
\affiliation{Department of Mathematics, Chinese Agriculture
University, Beijing, PR China}
\author{Yunbo Zeng}
\affiliation{ Department of Mathematical Sciences, Tsinghua
University, Beijing 100084, PR China}

\begin{abstract}
\textbf{Abstract}

We construct a new multi-component CKP hierarchy based on the
eigenfunction symmetry reduction. It contains two types of CKP
equation with self-consistent sources which Lax representations are
presented. Also it admits reductions to $k-$constrained CKP
hierarchy and to a (1+1)-dimensional soliton hierarchy with
self-consistent source, which include two types of Kaup-Kuperschmidt
equation with self-consistent sources and of bi-directional
Kaup-Kuperschmidt equation with self-consistent sources.
\end{abstract}
\maketitle

\textbf{1.Introduction}

Multi-component KP hierarchy attract a lot of interests from both
physical and mathematical points of view [1-8]. The multi-component
KP hierarchy given in [1] contains many physically relevant
nonlinear integrable systems such as Davey-Stewartson equation,
two-dimensional Toda lattice and three-wave resonant interaction
ones. Another kind of multi-component KP equation is the so-called
KP equation with self-consistent sources, which was initiated by
V.K. Mel'nikov [9-11].The first type of KP equation with
self-consistent sources (KPSCS) arises in some physical modes
describing the interaction of long and short wave [8-10,12], and the
second type of KPSCS is presented in [8,11,13]. Recently a method
was proposed in [8] to construct a new multi-component KP hierarchy
which includes first and second type of KPSCS. However, little
attention has been paid to the multi-component CKP hierarchy. In
addition, the CKP equation with self-consistent sources has not been
found out yet.

It is known that the Lax equation of KP hierarchy is given by [14]$$
L_{t_n }  = [B_n ,L] \eqno(1.1)$$where $$ L = \partial  + u_1
\partial ^{ - 1}  + u_2 \partial ^{ - 2}  + \cdots\eqno(1.2)$$is pseudo-differential
operator,$\partial$ denotes $ {\raise0.7ex\hbox{$\partial $}
\!\mathord{\left/
 {\vphantom {\partial  {\partial _x }}}\right.\kern-\nulldelimiterspace}
\!\lower0.7ex\hbox{${\partial _x }$}}$, $ u_i ,\;i = 1,2, \cdots$,
are functions in infinitely many variables $ t = (t_1 ,t_2 ,t_3 ,
\cdots )$ with $ t_1  = x $, and $ B_n  = L_ + ^n $ stands for the
differential part of $L^{n}$.

Owing to the commutativity of $\partial_{t_{n}}$ flows, we obtain
zero-curvature equations of KP hierarchy$$ B_{n,t_k }  - B_{k,t_n }
+ [B_n ,B_k ] = 0\eqno (1.3)$$ Eigenfunction $\Phi$ (adjoint
eigenfunction $\Phi^{\ast}$)satisfy the linear evolution equations
$$\Phi _{t_n }  = B_n (\Phi )\;(\Phi _{t_n }^ *   =  - B_n^ *  (\Phi ^
*  ))\eqno (1.4)$$ The compatibility condition of (1.4) is exactly (1.3).
\\The CKP hierarchy [15] is obtained from the KP hierarchy by
ignoring the time variables $ t_2 ,t_4 ,t_6 , \cdots $(i.e.
including only the odd time variables $ t_3 ,t_5 ,t_7 , \cdots $)and
by imposing at the same time the following antisymmetry condition on
the KP Lax operator$$L+L^{\ast}=0\eqno (1.5)$$ It follows
immediately from (1.5) that $$u_{2}=-\frac{1}{2}u_{1}^{'},
u_{4}=-\frac{3}{2}u_{3}^{'}+\frac{1}{4}u_{1}^{(3)},\ldots $$ and
$\Phi=\Phi^{\ast}, B_{n}=-B_{n}^{\ast}$ for $n$ odd. Taking $n=3,
k=5$, (1.3) and (1.5) lead to the CKP equation$$ u_{t _5 }-
\frac{5}{9}u_{t_3 }^{(2)}-\frac{5}{3}uu_{t_3}-\frac{5}{9}\partial
_x^{- 1}u_{t_3t_3 }+ \frac{1}{9}u^{(5)}+ \frac{25}{6}u^{'}u^{(2)}+
\frac{5}{3}uu^{(3)}- \frac{5}{3}u^{'}\partial_x^{- 1}u_{t_3 } +
5u^2u^{'}= 0 \eqno(1.6)$$where we use the notation
$u^{(i)}=\frac{\partial^{i}}{\partial x^{i}}$ in this paper.

In this paper, following the idea in [8] and using the eigenfunction
symmetry constraint, we firstly introduce a new type of Lax
equations which consist of the new time $\tau_{k}-$ flow and the
evolutions of wave functions. Under the evolutions of wave
functions, the commutativity of the evolutions of $\tau_{k}-$ flow
and $t_{n}-$ flow gives rise to a new multi-component CKP (mcCKP)
hierarchy. This hierarchy enables us to obtain the first and the
second types of CKP equation with self-consistent sources (CKPSCS)
and their Lax representations directly. This implies that the new
mcCKP hierarchy can be regarded as CKP hierarchy with
self-consistent sources (CKPHSCS). Moreover, this new mcCKP
hierarchy can be reduced to two integrable hierarchies: a
(1+1)-dimensional soliton hierarchy with self-consistent source and
the $k-$ constrained CKP hierarchy ($k-$ CKPH),which contain the
first type and the second type of Kaup-Kuperschmidt equation with
self-consistent sources and of bi-directional Kaup-Kuperschmidt
equation with self-consistent sources, respectively. Thus, the new
mcCKP hierarchy provides an effective way to find (1+1)-dimensional
and (2+1)-dimensional soliton equations with self-consistent sources
as well as their Lax representations. Our paper is organized as
follows. In section 2, we construct the new mcCKP hierarchy and show
that it contains the first and the second types of CKPSCS. In
section 3, the mcCKP hierarchy is reduced to a (1+1)-dimensional
soliton hierarchy with self-consistent source and the $k-$
constrained CKP hierarchy, respectively. In section 4, some
conclusions are given.

\textbf{2. New multi-component CKP hierarchy}

Following the idea in [8] and using the eigenfunction symmetry
constraint for CKP hierarchy [16], we define $\widetilde{B}_{k}$ by
$$ \tilde B_k  = B_k  + \sum\limits_{i = 1}^N {(q_i \partial ^{ - 1}
r_i  + r_i \partial ^{ - 1} q_i )}\eqno (2.1)$$ where $q_{i},r_{i}$
satisfy (1.4). Then we may introduce a new Lax equation given by
$$ L_{\tau _k }  = [B_k  + \sum\limits_{i = 1}^N {(q_i \partial ^{ - 1}
r_i  + r_i \partial ^{ - 1} q_i )} ,\;L]\eqno (2.2a)$$ $$ q_{i,t_n }
= B_n (q_i ),\;\;r_{i,t_n }  = B_n (r_i ),\;\,i = 1, \cdots ,N \eqno
(2.2b)$$ where $n, k$ are odd. \\\textbf{Lemma 1} $ [B_n ,r\partial
^{ - 1} q + q\partial ^{ - 1} r]_ -   = (r\partial ^{ - 1} q +
q\partial ^{ - 1} r)_{t_n }$ \\\textbf{ Proof:}  Set $B_n
=\sum\limits_{i = 1}^{n}a_{i}\partial^{i} $. Then we have $$
\begin{gathered} \ [B_n ,r\partial ^{ - 1} q + q\partial ^{ - 1} r]_
-= \sum\limits_{i = 1}^{n} {(a_{i} r^{(i)} \partial ^{ - 1} q +
a_{i} q^{(i)} \partial ^{ - 1} r) - \sum\limits_{i = 1}^{n}
{(r\partial ^{ - 1} q a_{i}\partial ^{i} + q\partial ^{ - 1} r a_{i}
\partial ^{i} )_ -  } } \\ = B_{n} (r)\partial ^{ - 1} q +
B_{n} (q)\partial ^{ - 1} r - \sum\limits_{i = 1}^{n}(r\partial ^{ -
1} q a_{i} \partial ^{i}  + q\partial ^{ - 1} r a_{i} \partial ^{i}
)_ -\end{gathered} $$
\\Applying integration by parts to the second term $$
  \sum\limits_{i = 1}^{n} {(r\partial ^{ - 1} q a_{i} \partial ^{i}  + q\partial ^{ - 1} ra_{i} \partial ^{i })_ -  }  =  \cdots  = \sum\limits_{i = 1}^{n} {( - 1)^{i} [r\partial ^{ - 1} (a_{i} q)^{(i)}  + q\partial ^{ - 1} (a_{i} r)^{(i)} ]}
  = r\partial ^{ - 1} B_n^ {*}  (q) + q\partial ^{ - 1} B_{n}^ {*}
  (r)$$ Noticing the facts that $ q^ *   = q, r^ *   = r, q_{t_n }^ *   =  - B_n^ *  (q^ *
  )$ and $r_{t_n }^ *   =  - B_n^ *  (r^ *  ) $, we can complete the
  proof immediately.\\\textbf{Theorem 1. }The commutativity of (1.1) and (2.2a) under (2.2b) leads to the following new integrable multi-component CKP (mcCKP)
  hierarchy  $$B_{n,\tau _k }  - (B_k  + \sum\limits_{i = 1}^N {(q_i \partial ^{ -
1} r_i  + r_i \partial ^{ - 1} q_i )} )_{t_n }  + [B_n ,B_k  +
\sum\limits_{i = 1}^N {(q_i \partial ^{ - 1} r_i  + r_i \partial ^{
- 1} q_i )} ] = 0 \eqno(2.3a)$$ \\ or equivalently $$
\begin{gathered}
  B_{n,\tau _k }  - B_{k,t_n }  + [B_n ,B_k ] + \sum\limits_{i = 1}^N {\{ [B_n ,r_i \partial ^{ - 1} q_i  + q_i \partial ^{ - 1} r_i ]}  - B_n (r_i )\partial ^{ - 1} q_i
\\- r_i \partial ^{ - 1} B_n (q_i ) - B_n (q_i )\partial ^{ - 1} r_i
- q_i \partial ^{ - 1} B_n (r_i )\}  = 0 \end{gathered}\eqno
(2.3a')$$ $$ q_{i,t_n }  = B_n (q_i ),\;r_{i,t_n }  = B_n (r_i
),\quad i = 1, \cdots ,N \eqno(2.3b)$$\\where $n$ and $k$ are
odd.Under (2.3b), the Lax pair for (2.3a) is given by $$ \psi _{t_n
}  = B_n (\psi ),\quad \psi _{\tau _k }  = [B_k  + \sum\limits_{i =
1}^N {(q_i \partial ^{ - 1} r_i  + r_i \partial ^{ - 1} q_i )}
](\psi )\eqno(2.4)$$ \\\textbf{Proof:}We will show that under
(2.3b), (1.1) and (2.2a) lead to (2.3a). For convenience, we assume
$N=1$ and denote $q_{1}, r_{1}$ by $q, r$. By (1.1), (2.2) and lemma
1, we have $$ \begin{gathered}
   B_{n,\tau _k }  = (L_{\tau _k }^n )_ +   = [B_k  + r\partial ^{ - 1} q + q\partial ^{ - 1} r,\;L^n ]_ +   = [B_k  + r\partial ^{ - 1} q + q\partial ^{ - 1} r,\;L_ + ^n ]_
   {+}+ [B_k  + r\partial ^{ - 1} q + q\partial ^{ - 1} r,\;L_ - ^n ]_ +
\\= [B_k  + r\partial ^{ - 1} q + q\partial ^{ - 1} r,\;L_ + ^n ] -
[B_k  + r\partial ^{ - 1} q + q\partial ^{ - 1} r,\;L_ + ^n ]_ -   +
   [B_k ,L_ - ^n ]_ +  \\ = [B_k  + r\partial ^{ - 1} q + q\partial ^{ - 1} r,\;B_n ] - [r\partial ^{ - 1} q + q\partial ^{ - 1} r,B_n ]_ -   + [B_n ,L^k ]_ +
    \\= [B_k  + r\partial ^{ - 1} q + q\partial ^{ - 1} r,\;B_n ] + (r\partial ^{ - 1} q + q\partial ^{ - 1} r)_{t_n }  + (B_k )_{t_n }
   \\= [B_k  + r\partial ^{ - 1} q + q\partial ^{ - 1} r,\;B_n ] + (B_k  + r\partial ^{ - 1} q + q\partial ^{ - 1} r)_{t_n }
   \end{gathered}$$ \\\textbf{Remark 1.} (2.3a') and (2.4) indicate that the mcCKP hierarchy can be regarded as the CKP hierarchy with self-consistent sources and is Lax integrable.
\\We now list some equations in this new mcCKP hierarchy.\\\textbf{Example 1 }(The first type of CKPSCS)
For $n=3, k=5$, (2.3)with $u=u_{1}$ leads to the first type of the
CKP equation with self-consistent sources $$ \begin{gathered}
u_{\tau _{5} }  - \frac{5} {9}u_{t_{3} }^{(2)}  - \frac{5}
{3}uu_{t_{3} } - \frac{5} {9}\partial _x^{ - 1} u_{t_{3} t_{3} }  +
\frac{1} {9}u^{(5)} + \frac{{25}} {6}u^{'} u^{(2)}  + \frac{5}
{3}uu^{(3)}  - \frac{5} {3}u^{'} \partial _{x}^{ - 1} u_{t_{3} }
  + 5u^2 u^{'}  + 2\sum\limits_{i = 1}^{N} {(q_{i}^{'} } r_{i } + q_{i} r_{i}^{'} ) =
  0,\\
  q_{i,t_{3} }  = q_{i}^{(3)}  + 3uq_{i}^{'}  + \frac{3}{2}u^{'} q_{i}
  ,\ \
  r_{i,t_{3 }}  = r_{i}^{(3)}  + 3ur_{i}^{'}  + \frac{3}{2}u^{'} r_{i },\ \  i = 1, \cdots ,N  \end{gathered} \eqno (2.5)$$
\\The Lax pair of (2.5) is given by $$ \begin{gathered}
\psi _{t _{3} }  = (\partial ^{3}  + 3u\partial  + \frac{3} {2}u^{'}
)(\psi ),\\
\psi _{\tau_{5} }  = (\partial ^{5 } + 5u\partial ^{3}  +
\frac{{15}} {2}u^{'} \partial ^{2}  + (\frac{5} {3}\partial _{x}^{ -
1} u_{t _{3} } + \frac{{35}} {6}u^{(2)}  + 5u^{2}  )\partial  +
[\frac{5} {6}u_{t _{3} }  + \frac{5} {3}u^{(3)}  + 5uu^{'}  +
\sum\limits_{i = 1}^{N} (q_{i} \partial ^{ - 1} r_{i}  + r_{i}
\partial ^{ - 1} q_{i} )]) (\psi )
\end{gathered} \eqno(2.6) $$ \\ \textbf{Example 2 (The second type of CKPSCS)}
For $n=5, k=3$, (2.3) with $u_{1}=u$ yields the second type of CKP
equation with self-consistent sources $$ \begin{gathered}
 u_{t_{5} }  - \frac{5}{9}u_{\tau _{3} }^{(2)}  - \frac{5} {3}uu_{\tau _{3} }  - \frac{5}
{9}\partial _{x}^{ - 1} u_{\tau _{3} \tau _{3} }  + \frac{1}
{9}u^{(5)} + \frac{{25}} {6}u^{'} u^{(2)}  + \frac{5} {3}uu^{(3)}  -
\frac{5} {3}u^{'} \partial _{x}^{ - 1} u_{\tau _{3} } + 5u^{2} u^{'}
= \\\frac{1} {3}\sum\limits_{i = 1}^{N} {[\frac{{10}} {3}(q_{i}
r_{i} )_{\tau _{3} }  + \frac{{20}} {3}q_{i}^{(3)} r_{i}  +
\frac{{20}} {3}r_{i}^{(3)} q_{i} + 10q_{i}^{(2)} r_{i}^{'}  +
10r_{i}^{(2)} q_{i}^{'} + 20uq_{i}^{'} } r_{i}  + 20uq_{i} r_{i}^{'}
+ 20u^{'} q_{i} r_{i }],\\ q_{i,t_{5} }  = q_{i}^{(5)}  +
5uq_{i}^{(3)}  + \frac{{15}} {2}u^{'} q_{i}^{(2)}  + (\frac{5}
{3}\partial _{x}^{ - 1} u_{\tau _{3} }  + \frac{{35}} {6}u^{(2)}  +
5u^{2}  + \frac{{10}} {3}\sum\limits_{i = 1}^{N} {q_{i} r_{i} }
)q_{i}^{'}  + [\frac{5} {6}u_{\tau _{3} }  + \frac{5} {3}u^{(3)}  +
5uu^{'}  + \frac{5} {3}\sum\limits_{i = 1}^{N} {(q_{i} r_{i} )^{'} }
]q_{i} ,
\\r_{i,t_{5} } = r_{i}^{(5)}  + 5ur_{i}^{(3)}  + \frac{{15}}
{2}u^{'} r_{i}^{(2)}  + (\frac{5} {3}\partial _{x}^{ - 1} u_{\tau
_{3} } + \frac{{35}} {6}u^{(2)}  + 5u^{2}  + \frac{{10}}
{3}\sum\limits_{i = 1}^{N} {q_{i} r_{i} } )r_{i}^{'}  +
 [\frac{5}{6}u_{\tau _{3} }  + \frac{5} {3}u^{(3)}  + 5uu^{'}  + \frac{5}
{3}\sum\limits_{i = 1}^{N} {(q_{i }r_{i })^{' }} ]r_{i },\\ i =
1,\cdots ,N \end{gathered} \eqno (2.7)$$ \\ The Lax pair of (2.7) is
given by $$ \begin{gathered}
 \psi _{\tau _{3} }  = [\partial ^{3}  + 3u\partial  + \frac{3}
{2}u^{'}  + \sum\limits_{i = 1}^{N} {(q_{i} } \partial ^{ - 1} r_{i}
+ r_{i}\partial ^{ - 1} q_{i} )](\psi ),\\
  \psi _{t_{5} }  =  (\partial ^{5}  + 5u\partial ^{3}  + \frac{15}
{2}u^{'} \partial ^{2}  + (\frac{5} {3}\partial _{x}^{- 1} u_{\tau
_3 } + \frac{{35}} {6}u^{(2)}  + 5u^{2}  + \frac{{10}}
{3}\sum\limits_{i = 1}^{N}{q_{i} r_{i} } )\partial  + [\frac{5}
{6}u_{\tau _{3} }  + \frac{5} {3}u^{(3)}  + 5uu^{'}  + \frac{5}
{3}\sum\limits_{i = 1}^{N} (q_{i} r_{i} )^{'}]) (\psi )
\end{gathered} \eqno (2.8)$$

\textbf{3. The $n-$ reduction and $k-$ constraint of (2.3)}

\textbf{3.1 The $n-$ reduction of (2.3)}
\\The $n-$ reduction of (2.3)is given by [14] $$L^{n}=B_{n}, \ \ or\ \  L^{n}_{-}=0 \eqno
(2.9)$$\\which implies that $$ L_{t_n }  = [B_n ,L] = [L^n ,L] = 0,\
\ B_{k,t_n }  = (L_ + ^k )_{t_n }  = 0,\ \ and \ \ \  q_{i,t_n }  =
r_{i,t_n }  = 0 \eqno (2.10)$$\\If $q_{i}$ and $r_{i}$ are wave
function, they have to satisfy [14] $$ B_n (q_i ) = L^n (q_i ) =
\lambda _i^n q_i ,\;B_n (r_i ) = L^n (r_i ) = \lambda _i^n r_i \eqno
(2.11)$$ \\So it is reasonable to impose the relation (2.11) in the
$n-$ reduction case. By using the Lemma 1 and (2.10), we can
conclude that the constraint (2.9) is invariant under the
$\tau_{k}-$ flow. Due to (2.10) and (2.11), one can drop $t_{n}-$
dependency from (2.3) and get the following (1+1)-dimensional
integrable hierarchy with self-consistent sources $$
\begin{gathered}
  B_{n,\tau _{k }}  + [B_{n} ,B_{k}  + \sum\limits_{i = 1}^{N} {(q_{i} \partial ^{ - 1} r_{i}  + r_{i} \partial ^{ - 1} q_{i} )} ] =
  0,\\ B_{n }(q_{i} ) = \lambda _{i}^{n} q_{i} ,\ \ B_{n} (r_{i} ) = \lambda _{i}^{n} r_{i} , \ \ i=1,\cdots, N \end{gathered} \eqno
  (2.12)$$\\with the Lax pair given by $$\begin{gathered} B_n (\psi ) = \lambda ^n \psi
  ,\\
   \psi _{\tau _k }  = [B_k  + \sum\limits_{i = 1}^N {(q_i \partial ^{ - 1} r_i  + r_i \partial ^{ - 1} q_i )} ](\psi ) \end{gathered} \eqno (2.13)$$

\textbf{Example 3 (The first type of KKESCS)} For $n=3, k=5$, (2.12)
presents the first type of Kaup-Kuperschmidt equation with
self-consistent sources $$ \begin{gathered}
 u_{\tau _{5} }  + \frac{1}
{9}u^{(5)}  + \frac{{25}} {6}u^{'} u^{(2)}  + \frac{5} {3}uu^{(3)} +
5u^{2} u^{'}  + 2\sum\limits_{i = 1}^{N} {(q_{i}^{'} } r_{i}  +
q_{i} r_{i}^{'} ) = 0, \\
q_{i}^{(3)}  + 3uq_{i}^{'}  + \frac{3} {2}u^{'} q_{i}  = \lambda
_{i}^{3} q_{i} \\
r_{i}^{(3)}  + 3ur_{i}^{'}  + \frac{3} {2}u^{'} r_{i}  = \lambda
_{i}^{3} q_{i} ,\ \ i = 1, \cdots ,N \end{gathered} \eqno (2.14)$$ \\
(2.13) with $n=3, k=5$ leads to the Lax pair of (2.14)$$
\begin{gathered}(\partial ^{3}  + 3u\partial  + \frac{3} {2}u^{'} )(\psi ) = \lambda
\psi , \\
\psi _{\tau _{5} }  = [\partial ^{5}  + 5u\partial ^{3}  +
\frac{{15}} {2}u^{'} \partial ^{2}  + (\frac{{35}} {6}u^{(2)}  +
5u^{2} )\partial  + (\frac{5} {3}u^{(3)}  + 5uu^{'} ) +
\sum\limits_{i = 1}^{N} {(q_{i} \partial ^{ - 1} r_{i}  + r_{i}
\partial ^{ - 1} q_{i} )} ](\psi ) \end{gathered} \eqno (2.15)$$\\
If we take $q_{i}=r_{i}=0$,  then (2.14) reduces to the
Kaup-Kuperschmidt equation [17]. \\\textbf{Example 4 (The first type
of BDKKESCS)} For $n=5, k=3$, (2.12) presents the first type of
bi-directional Kaup-Kuperschmidt equation with self-consistent
sources $$ \begin{gathered}
 - \frac{5}{9}u_{\tau _{3} }^{(2)}  - \frac{5} {3}uu_{\tau _{3} }  - \frac{5}
{9}\partial _{x}^{ - 1} u_{\tau _{3} \tau _{3} }  + \frac{1}
{9}u^{(5)} + \frac{{25}} {6}u^{'} u^{(2)}  + \frac{5} {3}uu^{(3)}  -
\frac{5} {3}u^{'} \partial _{x}^{ - 1} u_{\tau _{3} }\\
 + 5u^{2} u^{'}  = \frac{1}{3}\sum\limits_{i = 1}^{N} [\frac{{10}} {3}(q_{i} r_{i} )_{\tau _{3} }  +
\frac{{20}} {3}q_{i}^{(3)} r_{i}  + \frac{{20}} {3}r_{i}^{(3)} q_{i}
+ 10q_{i}^{(2)} r_{i}^{'}  + 10r_{i}^{(2)} q_{i}^{'}  + 20uq_{i}^{'}
 r_{i} +20uq_{i} r_{i}^{'}  + 20u^{'} q_{i} r_{i}],\\
 q_{i}^{(5)}  + 5uq_{i}^{(3)}  + \frac{{15}}
{2}u^{'} q_{i}^{(2)}  + (\frac{5} {3}\partial _{x}^{ - 1} u_{\tau
_{3 }} + \frac{{35}} {6}u^{(2)}  + 5u^{2}  + \frac{{10}}
{3}\sum\limits_{i = 1}^{N} q_{i} r_{i})q_{i}^{' } + [\frac{5}
{6}u_{\tau _{3} }  + \frac{5} {3}u^{(3)}  + 5uu^{'}  + \frac{5}
{3}\sum\limits_{i = 1}^{N}(q_{i} r_{i})^{'} ]q_{i}  =
\lambda _{i}^{5} q_{i} , \\
 r_{i}^{(5)}  + 5ur_{i}^{(3)}  + \frac{{15}}
{2}u^{'} r_{i}^{(2)}  + (\frac{5} {3}\partial _{x}^{ - 1} u_{\tau
_{3} } + \frac{{35}} {6}u^{(2)}  + 5u^{2}  + \frac{{10}}
{3}\sum\limits_{i = 1}^{N}q_{i} r_{i})r_{i}^{'}  +  \\
[\frac{5} {6}u_{\tau _{3} }  + \frac{5} {3}u^{(3)}  + 5uu^{'}  +
\frac{5} {3}\sum\limits_{i = 1}^{N} (q_{i} r_{i})^{'}]r_{i}  =
\lambda _{i}^{5} {r}_{i} ,i = 1, \cdots ,N \end{gathered}\eqno
(2.16)$$ \\ with the Lax pair given by $$ \begin{gathered} \psi
_{\tau _{3} }  = [\partial ^{3}  + 3u\partial  + \frac{3} {2}u^{'} +
\sum\limits_{i = 1}^{N} {(q_{i} } \partial ^{ - 1} r_{i}  + r_{i}
\partial ^{ - 1} q_{i} )](\psi ),\\
 \{ \partial ^{5}  + 5u\partial ^{3}  + \frac{{15}}
{2}u^{'} \partial ^{2}  + (\frac{5} {3}\partial _{x}^{ - 1} u_{\tau
_{3} } + \frac{{35}} {6}u^{(2)}  + 5u^{2}  + \frac{{10}}
{3}\sum\limits_{i = 1}^{N} {q_{i} r_{i} } )\partial  +\\
[\frac{5} {6}u_{\tau _{3} }  + \frac{5} {3}u^{(3)}  + 5uu^{'}  +
\frac{5} {3}\sum\limits_{i = 1}^{N} {(q_{i} r_{i} )^{'} } ]\} (\psi
) = \lambda ^5 \psi \end{gathered} \eqno (2.17)$$ \\If we take
$q_{i}=r_{i}=0$, then (2.16) reduces to the bi-directional
Kaup-Kuperschmidt equation [18,19].\\\textbf{3.2 The $k-$ constraint
of (2.3)}\\ The $k-$ constraint of (2.3)is given by [16] $$ L^k  =
B_k + \sum\limits_{i = 1}^N {(q_i \partial ^{ - 1} r_i  + r_i
\partial ^{ - 1} q_i )} \eqno (2.18)$$ It can seen that (2.18)
together with (2.2) lead to $L_{\tau_{k}}=0$ and $B_{n,\tau_{k}}=0$.
Then (2.3)becomes $k-$ constrained CKP hierarchy $$\begin{gathered}
(B_{k} + \sum\limits_{i = 1}^{N} {(q_{i} \partial ^{ - 1} r_{i}  +
r_{i}\partial ^{ - 1} q_{i} )} )_{t_{n} }  = [(B_{k}  +
\sum\limits_{i = 1}^{N} {(q_{i}\partial ^{ - 1} r_{i}  + r_{i}
\partial ^{ - 1} q_{i} ))_ {+} ^{\frac{n} {k}} } ,B_{k}  +
\sum\limits_{i = 1}^{N} {(q_{i} \partial ^{ - 1} r_{i}  + r_{i}
\partial ^{ - 1} q_{i} )} ],\\\ \
q_{i,t_{n} }  = (B_{k } + \sum\limits_{i = 1}^{N} {(q_{i} \partial
^{ - 1} r_{i}  + r_{i} \partial ^{ - 1} q_{i} ))_ {+} ^{\frac{n}
{k}} }(q_{i}) ,r_{i,t_{n }}  = (B_{k}  + \sum\limits_{i = 1}^{N}
{(q_{i}
\partial ^{ - 1} r_{i}  + r_{i} \partial ^{ - 1} q_{i} ))_ {+} ^{\frac{n}
{k}} }(r_{i}) ,\ \ i = 1, \cdots ,N \end{gathered}\eqno (2.19)$$
\\\textbf{Example 5 (The second type of KKESCS)} For $n=5,k=3$, (2.19) presents the second type of Kaup-Kuperschmidt equation with self-consistent sources
$$\begin{gathered}u_{t_{5} }  + \frac{1} {9}u^{(5)}  + \frac{{25}} {6}u^{'} u^{(2)}  +
\frac{5} {3}uu^{(3)}  + 5u^{2} u^{'}  = \frac{1} {3}\sum\limits_{i =
1}^{N} {[\frac{{20}} {3}q_{i}^{(3)} r_{i}  + \frac{{20}}
{3}r_{i}^{(3)} q_{i} } + 10q_{i}^{(2)} r_{i}^{'}  + 10r_{i}^{(2)}
q_{i}^{'}  + \\20uq_{i}^{'}
r_{i} + 20uq_{i} r_{i}^{'}  + 20u^{'} q_{i} r_{i} ],\\
q_{i,t_{5} }  = q_{i}^{(5)}  + 5uq_{i}^{(3)}  + \frac{{15}} {2}u^{'}
q_{i}^{(2)}  + (\frac{{35}} {6}u^{(2)}  + 5u^{2}  + \frac{{10}}
{3}\sum\limits_{i = 1}^{N }{q_{i} r_{i} } )q_{i}^{'}  + [\frac{5}
{3}u^{(3)}  + 5uu^{'}  + \frac{5} {3}\sum\limits_{i =
1}^{N} {(q_{i} r_{i} )^{'} } ]q_{i} ,\\
r_{i,t_{5} }  = r_{i}^{(5)}  + 5ur_{i}^{(3)}  + \frac{{15}} {2}u^{'}
r_{i}^{(2)}  + (\frac{{35}} {6}u^{(2)}  + 5u^{2}  + \frac{{10}}
{3}\sum\limits_{i = 1}^{N} {q_{i} r_{i} } )r_{i}^{'} +[\frac{5}
{3}u^{(3)}  + 5uu^{'}  + \frac{5} {3}\sum\limits_{i = 1}^{N} {(q_{i}
r_{i} )^{'} } ]r_{i} , \\i = 1, \cdots ,N
\end{gathered} \eqno (2.20)$$

\textbf{Example 6 (The second type of BDKKESCS)} For $n=3, k=5$,
(2.19) gives rise to the second type of bi-directional
Kaup-Kuperschmidt equation with self-consistent sources
$$\begin{gathered}
- \frac{5} {9}u_{t_{3} }^{(2)}  - \frac{5} {3}uu_{t_{3} }  -
\frac{5} {9}\partial _{x}^{ - 1} u_{t_{3} t_{3} }  + \frac{1}
{9}u^{(5)} + \frac{{25}} {6}u^{'} u^{(2)}  + \frac{5} {3}uu^{(3)}  -
\frac{5} {3}u^{'} \partial _{x}^{ - 1} u_{t_{3} }
 + 5u^{2} u^{'}  + 2\sum\limits_{i = 1}^{N} {(q_{i}^{'} } r_{i}  + q_{i} r_i^{'} ) = 0,
 \\
q_{i,t_{3} }  = q_{i}^{(3)}  + 3uq_{i}^{'}  + \frac{3} {2}u^{'}
q_{i} ,\ \ r_{i,t_{3} }  = r_{i}^{(3)}  + 3ur_{i}^{'}  + \frac{3}
{2}u^{'} r_{i} ,\ \ i = 1, \cdots ,N  \end{gathered}\eqno (2.21)$$

\textbf{4. Conclusion}

We firstly propose a new multi-component CKP hierarchy (mcCKP) based
on the eigenfuction symmetry constraint for the CKP hierarchy. This
mcCKP includes two types of CKP equation with self-consistent
sources. It admits reductions to the $k-$ constrained CKP hierarchy
containing the second type of some (1+1)-dimensional soliton
equation with self-consistent sources, and  reduction of CKP
hierarchy including the first type of some (1+1)-dimensional soliton
equation with self-consistent sources. Thus the mcCKP provides an
effective approach to find some (1+1)-dimensional and
(2+1)-dimensional soliton equations with self-consistent sources and
their related Lax representations. We notice that no solution has
been obtained not only for the first type of CKPSCS but for the
second type. So we will solve the integrable equations in the
forthcoming paper.

\textbf{Acknowledgment}

This work is supported by National Basic Research Program of China
(973 Program) (2007CB814800) and National Natural Science Foundation
of China (grant No. 10601028).


\begin{leftline}
\large\bf Reference\normalsize\rm
\end{leftline}

[1]  Date E , Jimbo M, Kashiwara M and Miwa T 1981 J. Phys. Soc.
Japan 50 3806-3812.

[2] Jimbo M and Miwa T 1983 Publ. Res. Inst. Math. Sci 19 943-1001.

[3] Sato M and Sato Y 1982 Soliton equations as dynamical systems on
infinite-dimensional Grassmann

~~~~~manifold. Nonlinear partial differential equations in applied
science (Tokyo).

[4]  Date E , Jimbo M, Kashiwara M and Miwa T 1982 Publ. Res. Inst.
Math. Sci 18 1077-1110.

[5]  V G Kac and J W van de Leur 2003 J. Math. Phys. 44 3245-3293.

[6]Johan van de Leur 1998 J. Math. Phys.39 2833-2847.

[7] Aratyn H, Nissimov E and Pacheva S 1998 Phys. Lett. A 244
245-255.

[8] Liu X J, Zeng Y B and Lin R L 2007 A new multi-component KP
hierarchy (Submitted)

[9]Mel'nikov V K 1983 Lett. Math. Phys. 7 129-136.

[10]Mel'nikov V K 1987Comm. Math. Phys. 112 639-652.

[11] Mel'nikov V K 1988 Phys. Lett. A 128 488-492.

[12] Xiao T and Zeng Y B, 2004 J. Phys. A: Math. Gen. 37 7143-7162.

[13]Wang H Y 2007 Some Studies on soliton equations with
self-consistent sources. PhD thesis,

~~~~Chinese Academy of Sciences.

[14]Dickey L A 2003 Soliton equation and Hamiltonian systems
(Singapore: World Scientific).

[15] Date E , Jimbo M, Kashiwara M and Miwa T 1981 J. Phys. Soc.
Japan 50 3813-3818.

[16] I. Loris 1999 Inverse Problem 15 1099-1109.

[17] D J Kaup 1980 Stud. Appl. Math. 62 189-216.

[18] Dye J M and Parker A 2001 J. Math. Phys. 42 2567-2589.

[19]Dye J M and Parker A 2002 J. Math. Phys.43 4921-4949.

\end{document}